# Tunable and temporally stable ferroelectric imprint through polarization coupling.

*Anirban Ghosh, Gertjan Koster\* and Guus Rijnders*

MESA+ Institute for Nanotechnology, University of Twente, P.O. Box 217, 7500AE Enschede, The Netherlands

## Abstract

Here we demonstrate a method to tune ferroelectric imprint, which is stable in time, based on the coupling between the non-switchable polarization of ZnO and switchable polarization of $PbZr_xTi_{(1-x)}O_3$. $SrRuO_3/PbZr_xTi_{(1-x)}O_3/ZnO/SrRuO_3$ heterostructures were grown with different ZnO thicknesses. It is shown that the coercive voltages and ferroelectric imprint varies linearly with the thickness of ZnO. It is also demonstrated that the ferroelectric imprint remains stable with electric field cycling and electric field stress assisted aging.

\* g.koster@utwente.nl

Ferroelectric devices based on metal-ferroelectric-metal (MFM) and metal-ferroelectric-semiconductor (MFS) heterostructures are interesting both from the application point of view e.g. in memories and other electronic applications as well as for understanding the fundamental physics[1-6]. Traditionally, the ferroelectric random access memory (FeRAM) and the ferroelectric field effect transistor (FeFET) have always been major fields of application[1-6]. Recently, a lot of focus is also being given to the ferroelectric resistive memory (RRAM) based on the ferroelectric tunnel junction (FTJ) and switchable diode effects because of their non-volatility, non-destructive and fast switching characteristics[7, 8]. It will also be of interest to see if the ferroelectric switching properties can be modified to our advantage for specific purposes.

In a FeFET the biggest challenge is having too short polarization retention times, which do not allow realistic applications[9-11]. In such a device the ferroelectric is in contact with a semiconductor, which has a finite screening length as compared to the Thomas-Fermi screening length in a metal) which gives rise to finite depolarizing field. Because of the depolarizing effect arising from this incomplete screening, the ferroelectric polarization is not stable for long time periods[10, 11]. It would be of great significance if one could make a FeFET in which the polarization is not lost by pinning it in one direction, thus stabilizing the ferroelectric polarization in one direction. This can be manifested by an imprint in the P-E hysteresis loop. This will lead to a longer retention time of the ferroelectric in the FeFET device, making it suitable for practical applications. Also, for newer generation kinetic memories e.g. phase change memory and RRAM the basic concept is to make two thermodynamically inequivalent temporally stable states with different (electrical) properties[8, 12]. Another application, in which ferroelectric imprint tuning is important is in the case of MEMS devices. It was shown recently that imprint in P-E hysteresis loop can be used advantageously to achieve a high figure of merit (figure of merit is inversely proportional to the dielectric permittivity of the ferroelectric at 0 V) in MEMS devices[13].

Ferroelectric imprint has been one of the main features of FeRAM devices[2, 3]. The asymmetry of the ferroelectric loop has been explained by several mechanisms, such as the presence of a ferroelectrically dead layer, a ferroelectric-electrode rectifying contact, by defect domain pinning or by interface charge injection. But none of these imprint phenomena have been shown to be temporally stable, nor have been demonstrated for tunability. Furthermore, the observed imprint appears to be very much processing dependent[14-18]. For all the above cases, it would be helpful if one could devise a method to create a ferroelectric imprint which is temporally as well as stable on multiple cycling, tunable and process independent.

In this paper we report on a simple method to tune the imprint of a ferroelectric which is temporally stable and in principle should be process independent. To achieve the goal of stabilizing one state of polarization over another, we have chosen heterostructures of ZnO and $PbZ_{0.58}Ti_{0.42}O_3$(PZT(58/42)) as a model system. PZT(58/42) in the rhombohedral phase (space group R3m) has a ferroelectric polarization along the $(111)_c$ direction (20). ZnO in its parent wurtzite structure (space group P6$_3$cm) has a non-switchable polarization along the $(0001)_h$ direction[19, 20]. Growing the PZT along the (pseudo) cubic $(111)_c$ direction provides us with the unique opportunity to study interaction between the switchable and non-switchable polarization at the interface. Because of the non-switchable polarization of ZnO one polarization state of PZT should be more stable than the other. The electric field due to the dipole is directly proportional to the thickness of the layer ($t_{ZnO}$) and its in-built polarization ($P_{ZnO}$). Hence, it would be very important to see how the ferroelectric imprint varies with the thickness of the ZnO. The thickness of the non-switchable ZnO layer is $d_{ZnO}$ and the thickness of the switchable PZT is $d_{PZT}$. The dielectric permittivity and polarization of ZnO and PZT are $\varepsilon_{ZnO}$, $\varepsilon_{PZT}$ and $P_{ZnO}$, $P_{PZT}$ respectively. Applying the condition of the continuity of the electric displacement field across the PZT/ZnO interface and calculate the applied voltage $V_{app}$ across the layer stack we can write

$$P_{PZT} + \varepsilon_0 \varepsilon_{PZT} E_{PZT} = P_{ZnO} + \varepsilon_0 \varepsilon_{ZnO} E_{ZnO} \quad (1)$$

$$V_{app} = E_{PZT} d_{PZT} + E_{ZnO} d_{ZnO} \quad (2)$$

$E_{ZnO}$, $E_{PZT}$ are the electric fields across the piezoelectric and the ferroelectric respectively and $V_{app}$ is the applied voltage. In order to find the offset voltage $V_{off}$, i.e. the difference in voltage that can produce the same amount of switching for both the biases we need to balance the $V_{PZT}$ for both the biases. The intrinsic coercive field of the PZT will remain the same only the applied voltage across the PZT will get changed for opposite biases due to the built in field. Following this we arrive at

$$V_{PZT} = V - V_{off} + d_{ZnO}(P_{PZT} + \varepsilon_0 \varepsilon_{PZT} E_{PZT})/\varepsilon_0 \varepsilon_{ZnO} \quad (3)$$

$$V_{off} = P_{ZnO} d_{ZnO} /\varepsilon_0 \varepsilon_{ZnO} \quad (4)$$

It can be seen from the Equation 4 that the imprint ($V_{off}$) is linearly dependent on the thickness of the non-switchable ZnO. In order to obtain a high imprint the non-switchable material should have a high polarization and a low dielectric constant. But on the other hand a low dielectric constant material will also increase the operating voltage, as is evident from Equation 3. We can also see from Equation 3. ($d_{ZnO}(P_{PZT} + \varepsilon_0 \varepsilon_{PZT} E_{PZT})/\varepsilon_0 \varepsilon_{ZnO}$) the presence of ZnO also leads to a symmetric increase (with respect to voltage) of coercive voltage with ZnO thickness. In our case the $d_{PZT}$ is fixed at 1μm, whereas $d_{ZnO}$ is varied from 25 nm to 150 nm. The presence of free charge carriers can in effect lead to the accumulation of free charge carriers at the PZT-ZnO interface. If, σ represents the trapped free charge carriers at the interface then

$$V_{off} = (P_{ZnO} + \sigma) \, d_{ZnO}/\varepsilon_0 \varepsilon_{ZnO} \quad (5)$$

It will in effect lead to a decrease of $V_{off}$[16, 21].

Herein, we show that we can tune the imprint in a ferroelectric by changing the thickness of the non-switching ZnO layer which is temporally stable. Capacitor structures of SrRuO$_3$(80 nm)/PZT(58/42)(1000 nm)/ZnO(25-150 nm)/SrRuO$_3$ (80 nm) were fabricated on STiO$_3$(111) substrates using pulsed laser deposition followed by photolithography and etching[22]. For the purpose of tuning the ferroelectric imprint devices with four different thicknesses of ZnO, 25 nm, 50 nm, 100 nm and 150 nm as well as a device without ZnO, were fabricated. Hereon, we name these samples PZT25, PZT50, PZT100, PZT150 and PZT0, respectively. The SrRuO$_3$ and PZT grow epitaxially on SrTiO$_3$ in the (111)$_c$ direction whereas ZnO grows preferentially in the (0001)$_h$ direction. The ferroelectric characterizations were carried out at room temperature using aixACCT 3000 TF Analyzer set up. From the reciprocal space map it was seen that the PZT films are fully strain relaxed.

In order, to measure the tuning of the ferroelectric hysteresis with ZnO thickness we measured the P-E hysteresis loop for the four different thicknesses of ZnO along with PZT without any ZnO. In Figure 1 we show the P-E hysteresis loops of the PZT0 and PZT100 devices measured at 100 Hz. It is observed that for the device without ZnO layer (PZT0) the saturation polarization ($P_s$) is around 34 μC/cm$^2$ and the coercive voltages are approximately ± 3 MV/m. In the case of the PZT100 sample the saturation polarization was 34 μC/cm$^2$ and the coercive voltages were -9.33 MV/m and 21.55 MV/m respectively for the negative and positive bias. In Figure 2. We plot the coercive fields for the opposite biases for all the samples the hysteresis loop opens up with increasing thickness of ZnO, and the coercive voltages also increase linearly with ZnO thickness according to Equation 3. To see the actual tunability of the imprint due to the fixed polarization of ZnO we plot in Figure 3 the offset voltage ($V_{off}$) (imprint ($V_{c+}+V_{c-}$)/2) (Equation 3). We can see that the $V_{off}$ is linearly proportional to the thickness of ZnO as in Equation 3. The coercive voltage was nearly independent of the measurement frequency (10 Hz-10 kHz)[3, 23]. Defect dynamics involving charging/discharging of the defect states result in large frequency dispersion of the hysteresis loops in ferroelectrics[3, 23, 24]. This shows that our electrical measurements are not dominated by defects and other relaxation mechanisms. This frequency dispersion study of our ferroelectric hysteresis loop shows our measurements are not dominated by artefacts resulting from leakage, and other space charge and other relaxation mechanisms and we are measuring the intrinsic switching characteristic of the system. The voltage offset per unit nm of ZnO is 0.05 V as determined from the slope of the imprint versus ZnO thickness plot. To ensure that the imprint effect is not dominated by the depletion effect because of the presence of two different materials across the PZT we measured the imprint for a 500 nm thick PZT with a 100 nm ZnO on top of it. In case the imprint is dominated by a depletion effect it should scale inversely with the thickness of PZT which was not observed (Figure 2 inset). Furthermore, symmetrical capacitance voltage measurements also showed there was no significant depletion layer formation inside the ZnO layer. To understand the effect of the free charge carriers in ZnO as described by Equation 5, we annealed all the samples at 600°C for 15 mins in $O_2$ atmosphere to decrease the free charge carriers in ZnO. We can see that upon annealing (Figure 3) the slope of the imprint vs ZnO thickness increases according to Equation 5. This shows that a more insulating ZnO layer would be more efficient for inducing imprint.

To make sure that the imprint is stable upon aging a cyclical voltage of ±35V was applied at 1 Hz frequency for up to 1000 secs. It was seen that even after 1000 secs of cycling the imprint didn't change by more than 4% for all the samples.

Waser *et al* [14, 15] showed that the imprint in ferroelectrics is caused by an interface field at the ferroelectric-electrode interface arising due to the finite separation between the polarization and screening charges. This interface field leads to charge injection from the electrode into the film which gets trapped at the interface and since the ferroelectric switching time is much faster than the charge detrapping time these trapped charges at the interface give rise to an imprint increasing with time. In the case when the imprint arises due to the interface field which is along the direction of polarization upon poling the sample in one direction the samples prefers that state due to electrical stress due to more charge injection and trapping[15, 18]. On application of a constant electric field along (or opposite to) the direction of this built in field would lead to an increase (decrease) of the imprint.

To test the role of the above mechanism role we investigated how the imprint evolves with time and also the effect of a constant voltage along to the direction of the imprint on the voltage shift. In Figure 4 we can see that for all the samples the imprint remains almost constant with time with which shows that the imprint is very stable to the effect of interface field. In Figure 4 inset we show that for PZT100 the application of a constant voltage leads to a very small change in the imprint as compared to the 0 V as a function of aging time. As a matter of fact, it's seen that the imprint actually decreases with time, indicating that the direction of the induced imprint is opposite to the direction of the interface field. Similar trends were also observed for all the PZT25 and PZT50 but for PZT0 the imprint increased with time as was also observed by Waser *et al*[14]. These observations show that the role of interface screening in inducing imprint is very small compared to that induced by the ZnO layer. Also, the imprint induced by interface field increase with increase in temperature which was not observed in our case, on the contrary the imprint increased with decrease in temperature (the results based on the temperature and frequency dependency of the hysteresis loop has been analysed in terms of nucleation and growth of domains and will be reported elsewhere).

In conclusion, we have shown that the imprint in a ferroelectric can be controlled and analyzed the underlying mechanism behind this phenomenon. Control of ferroelectric imprint is necessary for FeFET which has very short retention time as well as for RRAM based kinetic memories where it is necessary to make one state more stable than the other. Moreover, tunable imprint in a ferroelectric, which is also stable over time can lead to higher figures of merit and stable performance in energy harvesters. Our results have shown that the imprint or the offset voltage is directly proportional to the thickness of the ZnO layer thickness.

**Acknowledgements**

This work is supported by NanoNextNL, a micro and nanotechnology consortium of the Government of the Netherlands and 130 partners.

**References**

1. S. Mathews, R. Ramesh, T. Venkatesan and J. Benedetto, Science **276** (5310), 238-240 (1997).

2. C. A. Paz De Araujo, L. D. McMillan, B. M. Melnick, J. D. Cuchiaro and J. F. Scott, Ferroelectrics **104** (1), 241-256 (1990).

3. J. F. Scott, *Ferroelectric memories*. (Springer Science & Business Media, 2000).

4. W. Shockley, Proceedings of the IRE **40** (11), 1365-1376 (1952).

5. Y. Watanabe, Applied Physics Letters **66** (14), 1770 (1995).

6. S. Y. Wu, Ferroelectrics **11** (1), 379-383 (1976).

7. T. Choi, S. Lee, Y. Choi, V. Kiryukhin and S.-W. Cheong, Science **324** (5923), 63-66 (2009).

8. D. S. Jeong, R. Thomas, R. Katiyar, J. Scott, H. Kohlstedt, A. Petraru and C. S. Hwang, Reports on Progress in Physics **75** (7), 076502 (2012).

9. J. Hoffman, X. Pan, J. W. Reiner, F. J. Walker, J. P. Han, C. H. Ahn and T. P. Ma, Advanced materials **22** (26-27), 2957-2961 (2010).

10. T. Ma and J.-P. Han, Electron Device Letters, IEEE **23** (7), 386-388 (2002).

11. P. Wurfel and I. P. Batra, Physical Review B **8** (11), 5126-5133 (1973).

12. A. Chanthbouala, V. Garcia, R. O. Cherifi, K. Bouzehouane, S. Fusil, X. Moya, S. Xavier, H. Yamada, C. Deranlot, N. D. Mathur, M. Bibes, A. Barthelemy and J. Grollier, Nature materials **11** (10), 860-864 (2012).

13. S. Baek, J. Park, D. Kim, V. Aksyuk, R. Das, S. Bu, D. Felker, J. Lettieri, V. Vaithyanathan and S. Bharadwaja, Science **334** (6058), 958-961 (2011).

14. M. Grossmann, O. Lohse, D. Bolten, U. Boettger, T. Schneller and R. Waser, Journal of Applied Physics **92** (5), 2680 (2002).

15. M. Grossmann, O. Lohse, D. Bolten, U. Boettger and R. Waser, Journal of Applied Physics **92** (5), 2688 (2002).

16. A. K. Tagantsev and G. Gerra, Journal of Applied Physics **100** (5), 051607 (2006).

17. A. K. Tagantsev, I. Stolichnov, N. Setter and J. S. Cross, Journal of Applied Physics **96** (11), 6616 (2004).

18. W. L. Warren, B. A. Tuttle, D. Dimos, G. E. Pike, H. N. Al-Shareef, R. Ramesh and J. T. Evans, Japanese journal of applied physics **35** (2S), 1521 (1996).

19. V. M. Voora, T. Hofmann, M. Brandt, M. Lorenz, M. Grundmann, N. Ashkenov, H. Schmidt, N. Ianno and M. Schubert, Physical Review B **81** (19) (2010).

20. A. Janotti and C. G. Van de Walle, Reports on Progress in Physics **72** (12), 126501 (2009).

21. K. Abe, N. Yanase, T. Yasumoto and T. Kawakubo, Japanese Journal of Applied Physics **41** (Part 1, No. 10), 6065-6071 (2002).

22. M. D. Nguyen, M. Dekkers, E. Houwman, R. Steenwelle, X. Wan, A. Roelofs, T. Schmitz-Kempen and G. Rijnders, Applied Physics Letters **99** (25), 252904 (2011).

23. L. Pintilie and M. Alexe, Applied Physics Letters **87** (11), 112903 (2005).

24. L. Pintilie, *Charge transport in ferroelectric thin films*. (INTECH Open Access Publisher, 2011).

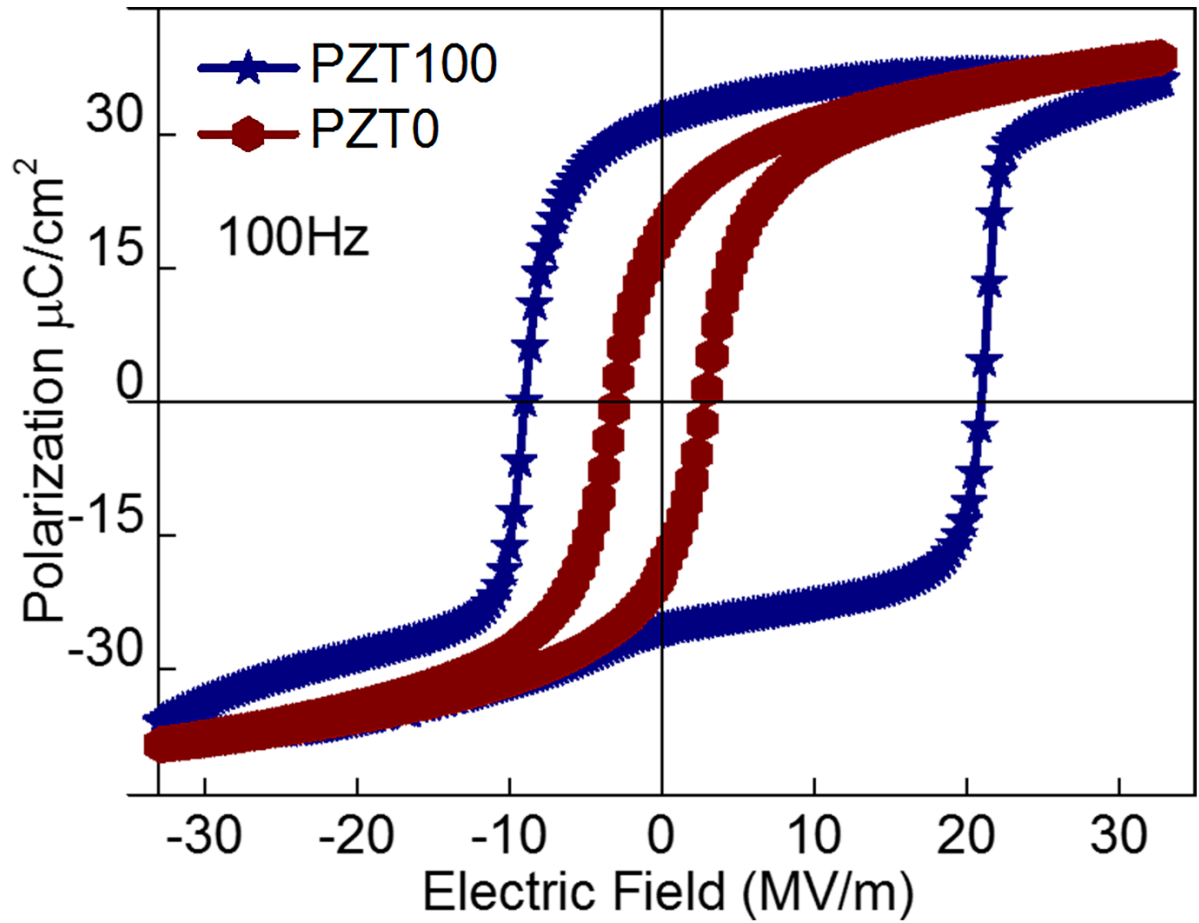

1. **Ferroelectric P-E hysteresis loop measured at 100 Hz of the PZT and PZT100.** The figure shows significant imprint for the PZT100 sample as well as opening up of the loop due to increase in the coercive fields.

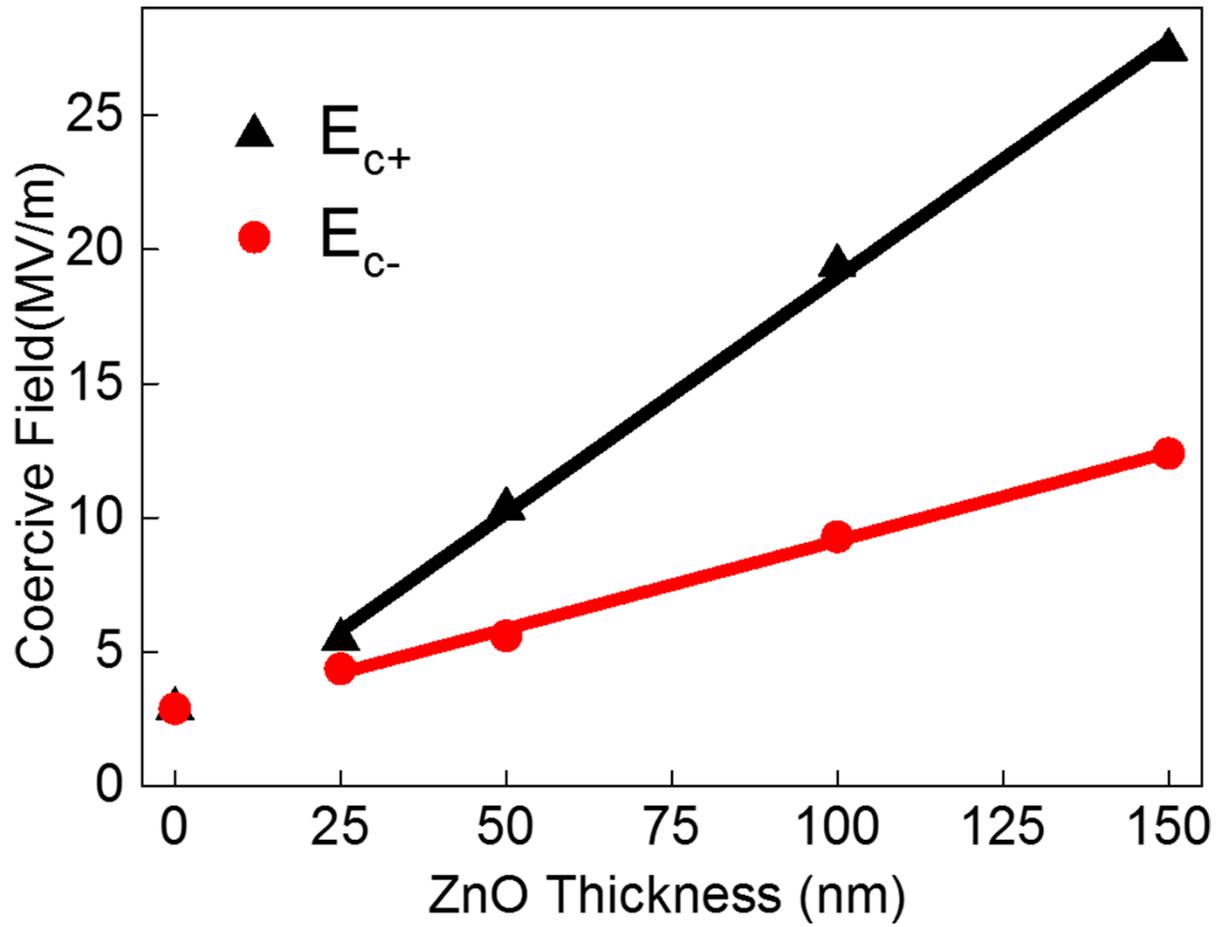

2. **Coercive field as function of ZnO thickness** This figure shows the linear variance of the coercive fields with the thickness of ZnO layer measured at 100Hz.

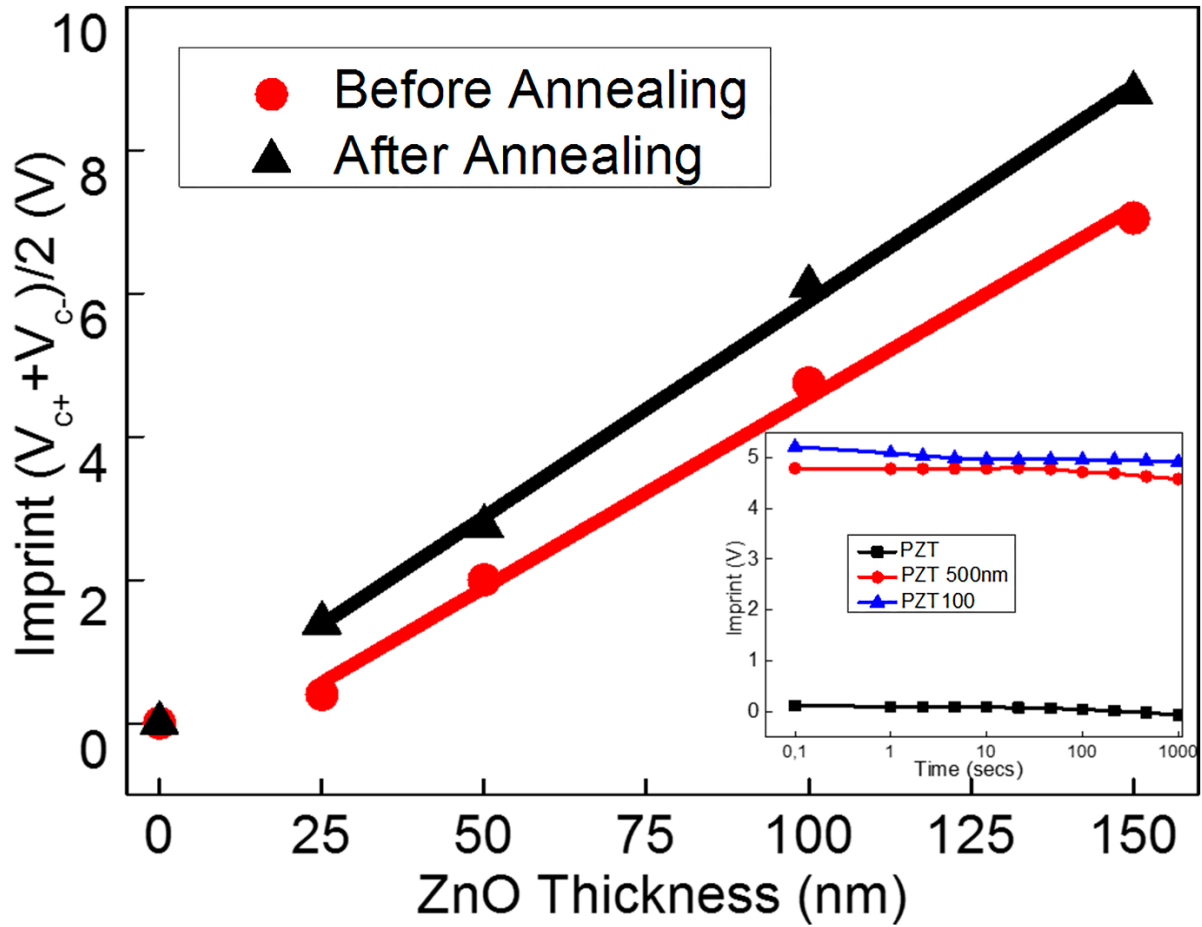

3. **Tuning of the imprint and with ZnO thickness and effect of annealing** a) Imprint (Offset voltage) of PZT (thickness fixed at 1 µm) samples measured as a function of ZnO thickness (25 nm, 50 nm, 100 nm, 150 nm). The figure shows that the imprint varies linearly with the thickness of ZnO. It can be seen upon annealing the imprint increases but still varies linearly. Inset shows the time dependent imprint behaviour of the PZT, PZT100 and PZT 500nm sample with 100 nm ZnO on top.

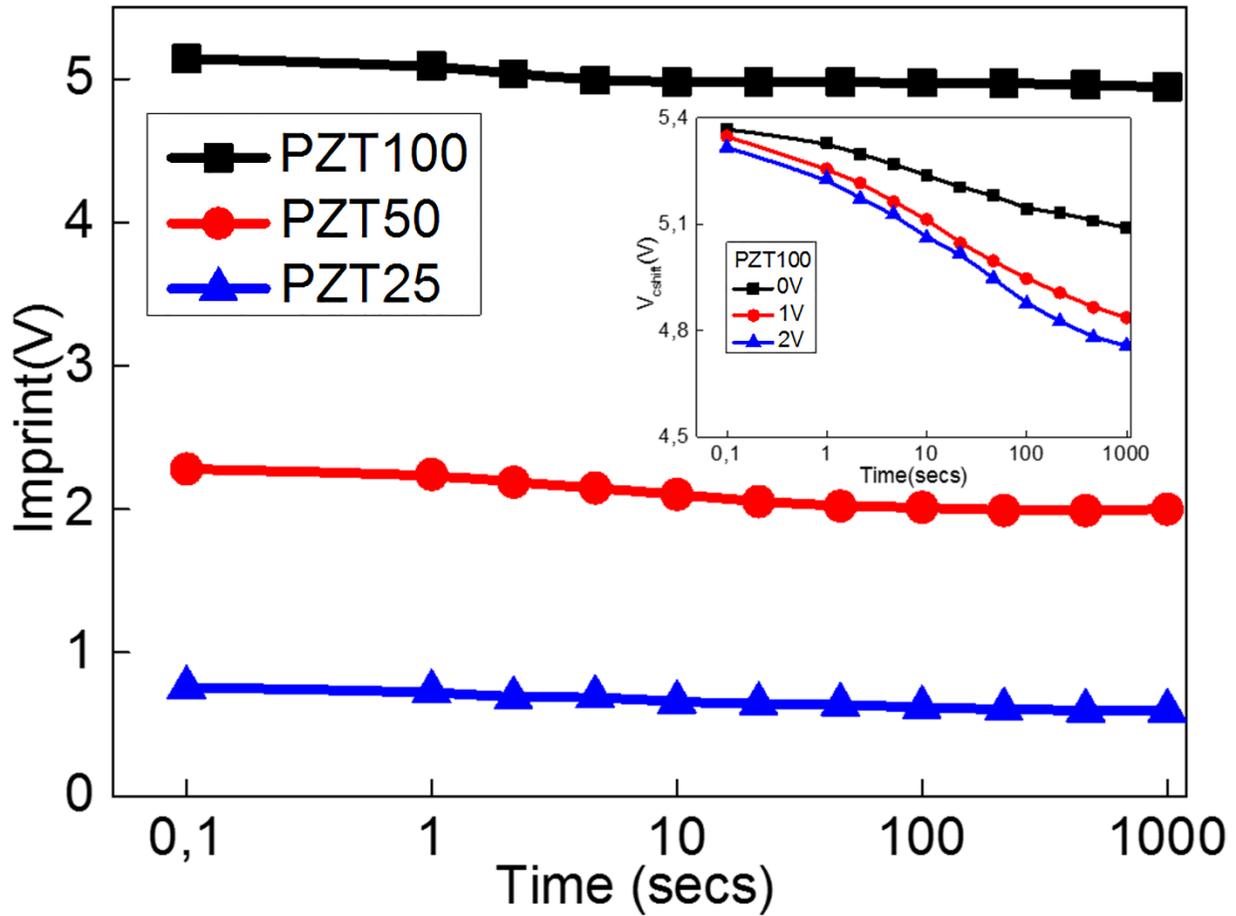

4. **Stability of Imprint with aging.** Stability of the imprint for all the samples with aging. Inset shows the evolution of the imprint for PZT100 upon application of a constant field along the direction of the imprint, the decrease in imprint with voltage shows that the direction of the imprint induced by the ZnO layer is opposite to that of the interface field.